\documentstyle[11pt]{article}
\def\ave#1{\left\langle #1\right\rangle}

\def\text{\rm}
\font\rm=cmr10 scaled\magstep1

\begin{document}
\begin{flushright}
\today
\end{flushright}
\begin{center}
\vspace{0.3in}
{\Large\bf Quantization of generic chaotic 3D billiard with smooth 
boundary I: energy level statistics }\\
\vspace{0.4in}
\large
Toma\v z Prosen
\footnote{e-mail: prosen@fiz.uni-lj.si}\\
\normalsize
\vspace{0.3in}
Physics Department, Faculty of Mathematics and Physics,\\
University of Ljubljana, Jadranska 19, 1000 Ljubljana, Slovenia\\
\vspace{0.3in}
\end{center}
\vspace{0.5in}

\noindent {\bf Abstract}
Numerical calculation and analysis of extremely high-lying energy spectra, 
containing thousands of levels with sequential quantum number up to 
62,000 per symmetry class, of a generic chaotic 3D quantum billiard is 
reported. The shape of the billiard is given by a simple and smooth 
de formation of a unit sphere which gives rise to (almost) 
fully chaotic classical dynamics. We present an analysis of (i) quantum 
length spectrum whose smooth part agrees with the 3D Weyl formula 
and whose oscillatory part is peaked around the 
periods of classical periodic orbits, 
(ii) nearest neighbor level spacing distribution and (iii) number 
variance.
 
Although the chaotic classical dynamics quickly and uniformly explores
almost entire energy shell, while the measure of the regular part of phase 
space is insignificantly small, we find small but significant deviations from 
GOE statistics which are explained in terms of  localization of 
eigenfunctions onto lower dimensional classically invariant manifolds.

\vspace{0.3 in}



\newpage
 
\noindent
In recent years a vast amount of numerical work has been invested in understanding the connection between classical chaotic dynamics and statistical properties of quantum energy spectra of autonomous Hamiltonian systems (see e.g. \cite{BTUPR} and references therein). 
Theoretical results relate the statistical properties of quantum energy levels either with the classical periodic orbits or with the eigenvalues of the statistical 
ensembles of random matrices \cite{GVZJ91}. In the later sense the strongest result is
numertically and theoretically supported
conjecture by Bohigas, Giannoni and Schmit \cite{BGS84} which identifies the 
asymptotic statistical properties of quantal energy spectra of classically fully chaotic 
systems with those of ensembles of Gaussian orthogonal/unitary/symplectic 
(GOE/GUE/GSE) random matrices; fine scale deviations from expected asymptotic, 
say GOE, statistics are related to localization --- non-uniformity of quantum eigenstates
on the energy surface.
So far, people have mostly studied systems with two degrees of freedom which are numerically much more accessible than higher dimensional systems, but which are also rather special since in 2D systems, unlike in higher dimensional systems, invariant tori (or cantori --- partial barriers) can geometrically divide the energy surfaces. However, numerics in quantum 3D systems is much more computer
time consuming than in 2D systems, so it is very difficult to
calculate highly excited 
eigenstates or long spectral stretches of chaotic 3D systems. 
Due to their energy-scaling property the cleanest systems to work with are certainly the 3D billiards ---
free particles moving freely in a bounded 3D domain and bouncing off
its boundary elastically. Recently,  Primack and Smilansky \cite{PS95} have calculated the first two thousand eigenstates of a 3D Sinai billiard, the free particle in the region inside a cube and outside a smaller concentric sphere.  Although the 3D Sinai billiard is classically fully chaotic --- ergodic, it is nongeneric in a sense that it possesses 2D and 3D invariant manifolds of non-isolated ``bouncing ball'' periodic orbits which greatly affect the properties of quantal spectra and produce significant deviations from GOE statistics for finite spectral samples as explained in \cite{PS95}.
\\\\
The first aim of this paper is to consider spectral statistics of a {\em generic} classically chaotic 3D system. We have chosen a 3D billiard with a smooth $C^\infty$ boundary. Since no such system is known to be rigorously ergodic we have defined a two-parameter family of 3D billiards whose shapes are given by simplest smooth deformations of a sphere:
The radial distance $r_B(\vec{n})$ from the origin to the boundary
is defined to be the following function of the 
direction $\vec{n},n^2=1$,
\begin{equation}
r_B(\vec{n}) = 1 + a (n_x^4 + n_y^4 + n_z^4) + b n_x^2 n_y^2 n_z^2
\label{eq:shape}
\end{equation}
which contains the two lowest order terms which preserve the cubic symmetry (the first two fully symmetric type ($\alpha$) cubic harmonics after Von Lage and Bethe \cite{VLB47}).
Using efficiently coded billiard's classical dynamics 
we have numerically explored the parameter space $(a,b)$ and found generic transition from integrable sphere, for $a=b=0$, via KAM scenario to more and more chaotic cases for increasing absolute values of parameters $a$ and $b$.
Finally we have chosen the shape (see figure 1): $a=-1/5,b=-12/5$ for which the billiard has the following properties:
(i) Almost all orbits of the system are uniformly and strongly chaotic 
with large average maximal liapunov exponent,
$\ave{\lambda_{\rm max}} = 0.54$,
and sharply peaked distribution of maximal finite-time liapunov exponets even for a short time of simulation per orbit (see figure 2).  There is no diffusion-like 
transport or partial barriers in phase space.  
(ii) The relative measure of the regular part of phase space $\rho_1\approx 10^{-3}$ is very small and should be almost insignificant for quantal calculations (see figure 2).  
(iii) The billiard is convex.
(iv) There are periodic orbits (for unspecified velocity they are (1+1)D 
invariant submanifolds)  which support all types of stability: besides the hyperbolic (including loxodromic) and a small fraction of elliptic orbits the system possesses also neutrally stable isolated parabolic orbits which touch the boundary only at the points of zero curvature radii,
$(\pm r_1,0,0),(0,\pm r_1,0),(0,0,\pm r_1), r_1=r_B(1,0,0)$.  
(v) There are (2+2)D invariant submanifolds in phase space $(\vec{r},\vec{p})$.
Besides the trivial symmetry planes $\{x=0,p_x=0\}$ and $\{x=y,p_x=p_y\}$ (and manifolds
 corresponding to cyclic permutations of  axes) and the boundary surface $\{r_B(\vec{r}/r) = r,\vec{p}\cdot\nabla (r_B(\vec{r}/r)-r)=0\}$, there is a nontrivial invariant submanifold $\{x+y=\pm z,p_x+p_y=\pm p_z\}$ (and its cyclic permutations) which is not a symmetry plane. The existence of such nontrivial flat invariant submanifold requires that on the curve, which is an intersection of the plane $x+y=\pm z$ and the boundary surface, any normal to the boundary surface should lie in the plane $x+y=\pm z$. For (\ref{eq:shape}) this is true if $b=12a$. The existence and construction of 
non-flat (2+2) invariant submanifolds is to the best of author's knowledge an open problem.
\\\\
The second aim of this paper is to demonstrate the power of the {\em
scaling method} for quantization of billiards, proposed recently by
Vergini and Saraceno \cite{VS95}, in {\em three dimensions}.  We have
slightly modified their method by using expansions in terms of
spherical waves $\phi_{klm}(r\vec{n}) = j_l(kr)Y_{lm}(\vec{n})$
instead of plane waves $\exp(i \vec{k}\cdot r\vec{n})$, the former
being much closer to billiard's geometry and naturally incorporating
the evanescent modes which are necessary to make the method {\em
arbitrarily accurate}.  The number of spherical waves necessary to
accurately quantize a 3D billiard for wavenumbers around $k$, 
$N_{\rm SW}(k) \approx (k r_{\rm max})^2$, where $r_{\rm max} = \max\{r_B(\vec{n});\vec{n}^2=1\}$, is almost optimal for shapes close to a sphere. The {\em efficiency} of the basis of spherical scaling functions $\{\phi_{klm}\}$ with respect to a 3D billiard with the shape (\ref{eq:shape}) is defined as 
$$ \eta = \frac{{\cal N}(k)}{{\cal N}_{\rm max}(k)} 
\approx \frac{V}{4\pi r_{\rm max}^3/3} $$ 
where ${\cal N}(k) = \#\{k_n < k\}  \approx V k^3/(6\pi^2)$ (see Weyl formula below (\ref{eq:weyl})) is the number of levels with wavenumber $k_n$ less than $k$ and 
$V=(1/3)\int d^2\vec{n} r_B^3(\vec{n})$ is the volume of the billiard, 
while ${\cal N}_{\rm max}(k) = 2 (k r_{\rm max})^3/(9\pi)$ is the number of levels less than $k$ in the smallest enscribed spherical billiard which has radius $r_{\rm max}$. 
For the spherical billiard efficiency $\eta$ is equal to one, 
whereas in order to have efficient method of quantization of a strongly chaotic system 
we have to compromise between largest possible efficiency $\eta$, on one hand, and largest 
possible average liapunov exponent $\ave{\lambda_{\rm max}}$ with small measure of  
regular orbits $\rho_1$, on the other hand.
We believe that our shape $a=-1/5,b=-12/5$, which has efficiency $\eta=0.866$, represents in this respect somehow an optimal choice.
More details on the scaling approach to quantization of 3D billiards can be found in a subsequent paper \cite{P96}.
\\\\
The billiard is preserved under 48-fold cubic symmetry group, so we have desymmetrized it and chosen only the class of singlet states belonging to fully symmetric irrep of a cubic group, 
i.e. we have considered a desymetrized part of 1/48 of a billiard bounded by the symmetry planes where von Neuman boundary conditions are imposed and the boundary surface where Dirichled boundary conditions are imposed. Desymmetrization also reduces the number of scaling functions, 
roughly by a factor of 48, 
which now become the {\em cubic waves} (CW): products of spherical Bessel functions
and {\em cubic harmonics} of type ($\alpha$) of Von Lage and Bethe \cite{VLB47}
(i.e. linear combinations of spherical harmonics $Y_{lm}$ with fixed $l$ 
which are invariant under the actions of the cubic group $O_h$),
$N_{\rm CW}(k) \approx (k r_{\rm max})^2/48$.
(See \cite{P96} for more details on the computation of CW.) 
We have calculated two spectral stretches containing around 7000 consecutive levels each: {\em sample I} contained 6973 consecutive levels on the interval $99.37 < k < 200.38$, and {\em sample II} contained 7133 consecutive levels on the interval $384.23 < k < 400.46$. In a single computer run, which required to solve a generalized eigenvalue problem \cite{VS95,P96} 
with matrices of size $N_{\text CW}(k)$, and has taken about 6 
hours of Convex $C3860$ CPU time we have obtained around 250 
accurate consecutive levels. ({\em Few tens} comparable computer jobs would be required to determine {\em a single} level with Heller's plane wave decomposition or boundary integral method.)
Maximal error was allowed to be $10^{-3}$ of the mean level spacing (MLS) 
but inaccuracy for most of eigenvalues was between $10^{-5}$ MLS and
$10^{-4}$ MLS.  It seemed that the number of accurate levels $N_{\rm converged}(k)$ per diagonalization is indeed proportional with the dimension of matrices 
$N_{\rm CW}(k)$ as claimed by Vergini and Saraceno \cite{VS95}, 
$$N_{\rm converged}(k) \approx 0.1 N_{\rm CW}(k) \approx 0.0016 k^2.$$
The scaling method has no problem of missing levels. 
We have compared the smooth part of the level counting function 
${\cal N}(E)$ with the generalization of the popular Weyl formula 
\cite{BH83} from 2D to 3D billiards
\begin{eqnarray}
{\cal N}_{\rm smooth}(k) &=& \frac{V}{6\pi^2}k^3 + \frac{A_N - A_D}{16\pi} k^2 + \label{eq:weyl} \\ &+& \frac{1}{12\pi^2}\left(\sum_j \frac{\pi^2-{\alpha^+_j}^2}{2\alpha^+_j} l^+_j - \sum_m \frac{\pi^2/2+{\alpha^-_m}^2}{2\alpha^-_m}l^-_m + C\right) k + {\cal O}(1)
\nonumber
\end{eqnarray}
where $V$ is the volume, $A_N$ ($A_D$) are the total areas of the boundary 
surfaces where Von Neuman (Dirichlet) boundary conditions are imposed.  
It is assumed that pairs of neighbouring 
smooth parts of the boundary surface with the same (opposite) type of boundary 
condition, DD or NN (DN or ND), intersect under the constant angle 
$\alpha^+_j$ ($\alpha^-_m$) along the curve of length $l^+_j$ ($l^-_m$). 
$C$ is the surface integral of the trace of curvature matrix
$$C =\!\!\!\int\limits_{\text boundary}\!\!\!d^2 S (\frac{1}{R_1} + 
\frac{1}{R_2}),$$ 
where $R_1$ and $R_2$ are the main curvature radii.
In figure 3 we plot the difference ${\cal N}(E)-{\cal N}_{\rm smooth}(E)$ 
for the sample I which locally average to zero so well that one could detect an error 
of $5\%$ in the curvature term $C$! Due to formula (\ref{eq:weyl}) the sample I contains the energy levels from $1,047$th to $8,018$th excited state, while the sample II contains energy levels from $54,706$th to $61,842$th excited state with uncertainty $\pm 2$.
\\\\
In order to analyze the relation of the spectra to classical periodic orbits we have calculated the 
quantum length spectrum (slightly modified w.r.t refs. \cite{SSCL93,PS95}) 
\begin{equation}
D(x) = \int\limits_{k_1}^{k_2} dk\; w(k;k_1,k_2) \cos(k x) d_{\rm osc}(k)
\label{eq:ls}
\end{equation}
where $d_{\rm osc}(k) = (d/dk)({\cal N}(k)-{\cal N}_{\rm smooth}(k))$ is an oscillatory part of the density of states and
$$w(k;k_1,k_2)=\frac{(k_2-k)(k-k_1)}{6(k_2-k_1)^3}$$ is a Welsch window function
which reduce oscillations due to final spectral length.
The length spectrum can be expressed semiclassically \cite{SSCL93} in terms of a sum of 
fat delta functions of width $\approx 1/(k_2-k_1)$ peaked around the periods of the classical 
periodic orbits (of a desimetrized billiard) with weights which are inversely 
proportional to the stability exponents.  
We have calculated $3692$ periodic orbits of a desymmetrized billiard for up to 12 
bounces and managed to identify most of the peaks in the length spectrum of sample I 
with the least unstable periodic orbits (figure 4a).  
Few unmatched peaks correspond to either missed or longer orbits.  
For the sample II the resolution $\delta x=2\pi/(k_2-k_1)$ is much worse and the broader peaks can no longer 
be associated with individual periodic orbits but with
the families of similar periodic orbits which have similar length and therefore can 
interfere constructively (figure 4b).
\\\\
To check the predictions of random matrix theory, the spectra $\{k_n\}$ were
unfolded $k_n\rightarrow e_n$ to unit MLS, $e_n = {\cal N}_{\rm smooth}(k_n)$.
First we have studied the nearest neighbor level spacing distribution (NNLSD) $P(S)$ 
of spacings $S_n = e_{n+1}-e_n$ and compared it with the predictions of GOE (accurate approximation of the exact infinitely dimensional formula for $P_{\rm GOE}(S)$ \cite{H91} have been used rather than the popular 2-dim Wigner surmise).
We have analyzed cumulative level spacing distribution 
$W(S) = \int_0^S ds P(s) = \#\{S_n < S\}/\#\{S_n\}$ in order to avoid losing information by bining and we have found small but statistically significant deviations of NNLSD from $P_{\rm GOE}(S)$ for both samples I and II (see figure 5).  
Deviations for the higher sample II, which are represented with uniform statistical error, $U(W_{\rm II}(S))-U(W_{\rm GOE}(S))$  (see figure 5b), where $U(W) = (2/\pi)\arccos\sqrt{1-W}$ (see \cite{PR93}), are functionally similar and only slightly smaller than deviations for the lower sample I, $U(W_{\rm I}(S))-U(W_{\rm GOE}(S))$.
(So deviations are clearly of systematic and not of statistical origin.)
We have tried to fit the NNLSD also with the phenomenological Brody distribution \cite{B73} which successfully describes the so-called fractional power law level repulsion \cite{PR93}, 
$P_{\beta}(S\rightarrow 0)\propto S^\beta$, 
typically associated with a localization of eigenstates.
We have also tried the two component Berry-Robnik NNLSD $P_{\rho_1}(S)$ \cite{BR84} which  assumes that the spectrum can be decomposed as statistically independent superposition of an uncorrelated level sequence with measure $\rho_1$ which may correspond to {\em localized} or {\em regular} states and GOE level sequence with measure $\rho_2=1-\rho_1$ which may correspond to {\em extended---delocalized chaotic} states.
As a result of the least square fit (see also figure 5b) we have obtained  for the sample I: $\beta_{\rm I}=0.872,\rho_{1\rm I}=0.024$ and (slightly more towards GOE) for the sample II: 
$\beta_{\rm II}=0.892,\rho_{1\rm II}=0.019$.
(The Brody fit is with both samples better than the Berry-Robnik fit, but neither of them
is statistically significant.) 
\\\\
The NNLSD depends mostly on the short range correlations between quantum energy levels. To compare the long range correlations with the predictions of random matrix theory we have studied 
the {\em number variance} 
$\Sigma^2(L) = \ave{(\tilde{\cal N}(e+L)-\tilde{\cal N}(e) - L)^2}$, 
the variance of the number of unfolded levels in the interval of
length $L$, where $\tilde{\cal N}(e) = \#\{e_n < e\}$
counts unfolded levels below $e$.
For the GOE spectra the number variance should increase logarithmically 
$\Sigma^2_{\rm GOE}(L)\approx \frac{2}{\pi^2}\log(2\pi L)$ whereas 
for the uncorrelated (Poissonian) sequence of levels the number variance is equal to the 
number itself $\Sigma^2_{\rm uncorr}(L) = L$. 
We have again found significant deviations from GOE statistics for both spectral 
samples but deviations for the lower sample I were significantly larger than for the higher sample II which is consistent with the validity of GOE statistics in the ultimate semiclassical limit. 
We conclude that localized 
states constitute approximately independent uncorrelated spectral
subsequence, with respect to {\em long-range} correlations, since the number variance is excellently captured by the ansatz of Seligman and Verbaarschot \cite{SV85} (see figure 6) 
\begin{equation}
\Sigma^2(L) = \tilde{\rho}_1 L + \Sigma^2_{\rm GOE}(\tilde{\rho}_2 L),
\label{eq:sig}
\end{equation} 
with the following measures of uncorrelated (localized) subsequence 
$\tilde{\rho}_{1 I}=0.072$ and $\tilde{\rho}_{1 II}=0.028$, for the samples I and II, respectively. 
In other words: in the samples I and II there should be $\approx 500$ and 
$\approx 200$ strongly localized  eigenstates (scars), respectively  \cite{P96}.  
However, note that one should not expect consistency with Berry-Robnik parameter 
$\rho_1$  (as determined from non-significant fit by the Berry-Robnik distribution 
\cite{BR84}), since the two spectral subsequences are expected to be {\em short-range} 
correlated: the localized and extended wavefunctions typically do
overlap giving rise to the repulsion of nearby levels.
The formula (\ref{eq:sig}) is expected to be valid only in the universal regime 
$L \ll L^*$, where $L^*$ is a saturation length scale \cite{B89} 
$L^* \approx \frac{1}{l^*}(d/dk){\cal N}(k)$ and $l^*$ is the length of the shortest periodic orbit of the desymmetrized billiard, while for $L\approx L^*$ the shortest periodic orbits of the system start to govern the behavior of a number variance in a system-dependent non-universal way \cite{B89,AS95}. 
Note that $L^* \propto  k^{(D-1)} \propto {\cal N}^{(D-1)/D}$, where 
$D$ is the number of freedoms, so the universality region of $L$ is much 
larger in 3D systems than in 2D systems
for comparable sequential quantum numbers: so the quantitative effect of quantum
eigenstates which are enhanced in the neighborhood of classically invariant
sets  (scars) on the long-range energy level statistics is cleaner in higher 
dimensions.
\\\\
In a conclusion we should stress that using an efficient numerical technique \cite{VS95} 
(see also \cite{P96}) we have been able to calculate high-quality energy spectra of  a 
generic 3D chaotic billiard going up to $62,000$th  excited state within fixed symmetry 
class. Using sensitive tests on various energy level statistics, such as short-range nearest
neighbor level spacing distribution and long-range number variance, 
we have detected small but significant deviations from the GOE statistics 
which is an indication of the localization of  eigenstates, 
although the classical dynamics is non-diffusive and strongly chaotic.  
It has been shown that the spectrum can be decomposed 
with respect to long-range correlations to a GOE subsequence of extended 
chaotic states and an uncorrelated (Poisson) subsequence  of  strongly localized states.  
The structure of individual eigenfunctions a chaotic 3D billiard (\ref{eq:shape}) 
in various representations, and examples of strongly localized states, 
are studied in \cite{P96}.
The present study also motivates further study of energy level statistics in a generic
classically mixed regime, which is also realized by our billiard with smaller absolute 
values of parameters $a$ and $b$, to test the Berry-Robnik surmise in 3D, and 
to search for quantal consequences of genuine 3D aspects of dynamics, such as Arnold's 
diffusion and its effect on quantum energy level statistics.

\section*{Acknowledgments}
Enlightening discussions with H.Primack, E.Vergini and M.Saraceno as well as financial support of the Ministry of Science and Technology of the Republic of Slovenia are gratefully acknowledged.

\section*{Figure captions}
\noindent {\bf Figure 1:}
The shape of the boundary of the
chaotic 3D billiard (\ref{eq:shape}) for $a=-1/5,b=-12/5$.

\bigskip
\noindent {\bf Figure 2:}
The cumulative distributions of (finite-time) maximal liapunov exponents are plotted for
7 different lengths of orbits (with unit velocity)
$l=t= 400, 400\sqrt{2}, 800, 800\sqrt{2}, 1600, 1600\sqrt{2},3200$.
Each distribution was generated using $10000$ orbits with 
initial conditions chosen randomly and uniformly over the phase-space. All
distributions have almost the same average maximal liapunov exponent
$\ave{\lambda_{\rm max}}\approx 0.540$ with small and decreasing dispersions
$\ave{(\lambda_{\rm max} - \ave{\lambda_{\rm max}})^2}=
 0.0567,0.0493,0.0437,0.0391,0.0352,0.0319,0.0290$, respectively.
The region of small liapunov exponents, indicated with a dotted line, is magnified in 
the inset. The size of a
jump of a cumulative distribution of long-time liapunov exponents for small values of 
the exponent (which is around one tick of the inset $= 10^{-3}$) is an estimate (or 
an upper bound) of the relative volume of the regular part of phase space $\rho_1$.
 \bigskip

\noindent {\bf Figure 3:}
The oscillatory part of the level counting function,
${\cal N}(k) - {\cal N}_{\rm smooth}(k)$, is shown for the sample $I$. 
Local average of the noisy data over 1000 neighboring levels is 
shown as the small band whose width is an estimated statistical error.
The fact that the average band overlaps with or slightly fluctuates  
(oscillations are due to shortest periodic orbits) 
around abscissa on a scale of few percents of
a level is a very good `experimental' test of 3D Weyl formula 
(\ref{eq:weyl}). 

\bigskip
\noindent {\bf Figure 4:}
The quantum
length spectrum $D(x)$ (\ref{eq:ls}) is shown for sample $I$ (a) and
sample $II$ (b).
Vertical dotted lines indicate the lengths of least unstable
(with $\sqrt{|\det{M-1}|} < 100$, where $M$ is a monodromy matrix) and shortest 
(up to 12 bounces) periodic orbits of  a desymmetrized billiard which match with 
peaks of quantum length spectrum of the sample I.
For the sample I (a) we have better resolution  that for the sample II (b).
The pattern of dots denotes the type of stability: $.\,\,\,.\,\,\,.$ elliptic orbit
(two pairs of  unimodular eigenvalues of $M$),
$..\,\,..\,\,..$ hyperbolic orbit (one pair of  unimodular and one pair of real
eigenvalues of $M$), 
$...\,...\,...$ doubly hyperbolic orbit (two pairs of real eigenvalues of  $M$),
and $............$ loxodromic orbit (four complex eigenvalues of $M$).

\bigskip
\noindent {\bf Figure 5:}
The cumulative NNLSD $W(S)$ is shown in (a) for sample I (thin curve) and for
sample II (thick curve). In the inset we show magnified small $S$ region.
The dotted curves are the GOE and Poissonian NNLSD.
Below (b) we show the deviations from GOE statistics with a uniform error and
with uniform density of points along abscissa, $U(W(S))-U(W_{\rm GOE}(S))$
against $W(S)$, where $U(W)=(2/\pi)\arccos\sqrt{1-W}$.
The thin and thick noisy curve are the data for samples I and II, respectively,
the dotted and dashed curves are the best fitting Brody and
Berry-Robnik distributions, respectively, the thin ones for the sample I and the thick 
ones for the sample II. The horizontal dotted lines indicate $\pm$ one-sigma band of 
estimated statistical error for the numerical data. 

\bigskip 
\noindent {\bf Figure 6:}
We show the number variance $\Sigma^2(L)$ for sample I (thin curve) and sample II
(thick curve). The dashed curves are the corresponding best  fits to 
Seligman-Verbaarschot formula (\ref{eq:sig})  (see text), which has been fitted on
the interval $2 \le L \le 12$ for the sample $I$ and on the interval $2 \le L \le 40$ 
for the sample $II$.
One should observe excellent agreement with the model (dashed) curves
(\ref{eq:sig}) for the spectral sample I up to $L\approx 12$, and for the spectral
sample $II$ up to $L\approx 40$ where notable
agreement is continued up to
$L\approx 100$.
The dotted curve is the number variance for GOE.

\begin{thebibliography}{99}
\bibitem[$1$]{BTUPR} {\sc O. Bohigas, S. Tomsovic and D. Ullmo},
Phys. Rep. {\bf 223}, 4 (1993);
{\sc T. Prosen and M. Robnik}, J. Phys. A {\bf 27}, 8059 (1994);
{\sc T. Prosen} Physica D{\bf 91}, 244 (1996).
\bibitem[$2$]{GVZJ91} {\sc M.-J. Giannoni, A. Voros and J. Zinn-Justin}, 
eds., {\em Chaos and Quantum Physics}, Proceedings of the 1989 Les Houches 
Summer School, (Elsevier Science Publishers B.V., Amsterdam, 1991).
\bibitem[$3$]{BGS84} {\sc O. Bohigas, M.-J. Giannoni and C. Schmit},
Phys. Rev. Lett. {\bf 52}, 1 (1984).
\bibitem[$4$]{PS95} {\sc H. Primack and U. Smilansky}, Phys. Rev. Lett.
{\bf 74}, 4831 (1995).
\bibitem[$5$]{VLB47} {\sc F.C. Von der Lage and H.A. Bethe}, Phys. Rev. {\bf 71}, 
612 (1947)., 4831 (1995).
\bibitem[$6$]{VS95} {\sc E. Vergini and M. Saraceno}, 
Phys. Rev. E {\bf 52}, 2204 (1995).
\bibitem[$7$]{P96} {\sc T. Prosen}, ``Quantization of generic chaotic 3D billiard
with smooth boundary II: structure of high lying eigenstates'', 
{\em Preprint}, submitted to Phys. Lett. A.
\bibitem[$8$]{BH83} {\sc H.P. Baltes and E.R. Hilf}, {\em Spectra of Finite Systems},
(Bibliographysches Institut, Mannheim, 1976)
\bibitem[$9$]{SSCL93} {\sc M. Sieber, U. Smilansky, S.C. Creagh and R.G.
Littlejohn}, J. Phys. A {\bf 26}, 6217 (1993).
\bibitem[$10$]{H91} {\sc F. Haake}, {\em Quantum Signatures of Chaos}, 
(Springer-Verlag Berlin Heidelberg 1991).
\bibitem[$11$]{PR93} {\sc T. Prosen and M. Robnik}, J. Phys. A {\bf 26}, 2371 (1993).   
\bibitem[$12$]{B73} {\sc T. A. Brody}, Lett. Nuovo Cimento {\bf 7}, 482 (1973).
\bibitem[$13$]{BR84} {\sc M.V. Berry and M. Robnik}, J. Phys. A {\bf 17}, 2413 (1984).
\bibitem[$14$]{SV85} {\sc T.H. Seligman and J.J.M. Verbaarschot}, 
J. Phys. A {\bf 18}, 2227 (1985).
\bibitem[$15$]{B89} {\sc M.V. Berry},  {\em Some Quantum to Classical Asymptotics} 
in ref.\cite{GVZJ91}, p.251.
\bibitem[$16$]{AS95} {\sc R. Aurich and F. Steiner}, Physica D{\bf 82}, 266 (1995).
\end{thebibliography}
\end{document}